\documentclass[aip,rsi,reprint,graphicx]{revtex4-1} 
\usepackage{graphicx}
\usepackage{dcolumn}
\usepackage{bm}
\usepackage[mathlines]{lineno}
\usepackage{amsmath}
\usepackage[normalem]{ulem} 
\usepackage{xcolor}
\usepackage[compact]{titlesec}
\titlespacing{\section}{0pt}{2ex}{1ex}
\titlespacing{\subsection}{0pt}{1ex}{0ex}
\titlespacing{\subsubsection}{0pt}{0.5ex}{0ex}

\begin{document}
\setlength{\belowdisplayskip}{1pt} \setlength{\belowdisplayshortskip}{1pt}
\setlength{\abovedisplayskip}{1pt} \setlength{\abovedisplayshortskip}{1pt}

\title[]{Time-lens Photon Doppler Velocimetry (TL-PDV)}

\author{Pinghan Chu
}\email{pchu@lanl.gov} \thanks{ correspondence co-first author}
\affiliation{ 
Los Alamos National Laboratory, Los Alamos, New Mexico 87545, USA
}%
\author{Velat Kilic}\email{velat\_kilic@jhu.edu} \thanks{co-first author}
\author{Mark A. Foster}\email{mark.foster@jhu.edu}
\affiliation{
Johns Hopkins University, Baltimore, MD 21218 USA\\
}
\author{Zhehui Wang}\email{zwang@lanl.gov}
\affiliation{ 
Los Alamos National Laboratory, Los Alamos, New Mexico 87545, USA
}%
\date{\today}
\begin{abstract}
We describe a time lens to expand the dynamic range of photon Doppler velocimetry (PDV) systems. The principle and preliminary design of a time-lens PDV (TL-PDV) are explained and shown to be feasible through simulations. In a PDV system, an interferometer is used for measuring frequency shifts due to the Doppler effect from the target motion. However, the sampling rate of the electronics could limit the velocity range of a PDV system. A four-wave-mixing (FWM) time lens applies a quadratic temporal phase to an optical signal within a nonlinear FWM medium (such as an integrated photonic waveguide or highly nonlinear optical fiber). By spectrally isolating the mixing product, termed the idler, and with appropriate lengths of dispersion prior and after to this FWM time lens, a temporally magnified version of the input signal is generated. Therefore, the frequency shifts of PDV can be ``slowed down" with the magnification factor $M$ of the time lens. $M=1$ corresponds to a regular PDV without a TL. $M=10$ has been shown to be feasible for a TL-PDV. Use of this effect for PDV can expand the velocity measurement range and allow the use of lower bandwidth electronics. TL-PDV will open up new avenues for various dynamic materials experiments. 
\end{abstract}
\maketitle
\section{Introduction}
\label{sec:introduction}
Velocimetry is a measurement technique for velocity of motion that plays an important role in the study of shock physics. Doppler shifted optical interferometry using lasers has been applied to velocimetry shortly after the emergence of the laser. Several techniques have been developed. In the 1970s, velocity interferometer system for any reflector (VISAR)~\cite{Barker:1972} was developed to measure the velocity-dependent phase changes via a spatial fringe pattern intensity by interfering the reflected laser light from a moving surface with unmodulated laser. In the 1980s, the Fabry-Perot interferometer (FPI)~\cite{McMillan:1988} was also developed which relies on the fringe position. VISAR can record data for a longer time compared to the FPI and is more sensitive while FPI can measure multiple discrete velocities simultaneously. More recently, a concept based on the ratio of the light intensities from two detectors, correlated-intensity velocimeter for arbitrary reflector (CIVAR)~\cite{Wang:2006}, has been proposed but so far has no further development. However, due to its simplicity and robustness under realistic experimental conditions~\cite{Dolan:2010}, the photon Doppler velocimetry (PDV)~\cite{Strand:2006} has become more popular by measuring frequency shift and avoiding fluctuations in the light intensity as the one in VISAR and FPI.

In PDV, the optical frequency shift of a laser reflected from a moving surface due to the Doppler effect is used to measure the velocity of the object. The light from a laser of the frequency $f_0$ is launched onto a moving surface and the frequency ($f_d$) of the reflected light is generated. In a PDV system, the reflected light and non-Doppler-shifted light from the original laser are mixed to generate a beat signal that is measured using a photodetector. Current PDV systems can measure a broad range of velocities from a few m/s to roughly 50 km/s, with a high accuracy of 10 m/s and time resolution of 10 ns, which can be recorded using digitizers with 10s of GHz sampling frequency~\cite{Moro:2014}. Although electrical bandwidth has been a limiting factor for maximum detectable velocities, various approaches have been demonstrated to increase the velocity dynamic range~\cite{Dolan:2020}. In one example, high-frequency signals resulting from high velocities can be converted to lower-frequency signals within the detection band by mixing with a local oscillator. However such approaches can create spectral artifacts and limit the low end of the detectable velocity range. In another example, the leapfrog techniques using multiple laser wavelengths can extend the velocity range of PDV with a significant increase in complexity and cost. Most recently, time-stretched PDV has been demonstrated~\cite{Mance:2019} by creating several time stretched replicas of the input waveform. However this again comes with a significant increase in optical hardware complexity and limits sensitivity and the ability for continuous time coverage. 

Here we describe the use of temporal imaging with a time lens to expand the velocity dynamic range of PDV. The principle of the time lens can be understood by using the space-time duality of electromagnetic waves~\cite{Kolner:1994}. Analogous to a spatial lens that can serve as an image magnifier in imaging systems when combined with free-space propagation (diffraction), a time lens can be used as a time magnifier in a dispersive medium. A time lens is created by imparting a quadratic temporal phase onto a temporal optical waveform much the same as spatial lens imparts a quadratic spatial phase onto a spatial optical signal. Thus a time lens can be used to magnify, compress, and Fourier transform a temporal optical waveform~\cite{Kolner:1989}. For temporal imaging systems,  dispersive propagation is generally achieved using optical fibers~\cite{Haus:1984}. Additionally, several distinct methods can be used to generate the temporal quadratic phase for the time lens, such as an electro-optical phase modulator driven with a sinusoidal voltage~\cite{Kauffman:1994} or nonlinear optical processes~\cite{Mouradian:2000,Salem:2008}. One of the key properties of the time lens is the maximum achievable phase shift. For electro-optical implementations, the maximum phase shift is limited by the maximum tolerable voltage of the phase modulator. For nonlinear optical cross phase modulation, this is limited by the pump powers and distortions to the pump pulse from self-phase modulation. The largest achievable phase shifts are possible using nonlinear optical wave mixing approaches with magnification factors larger than 500, which has been demonstrated using four-wave mixing (FWM) in a silicon photonic waveguide chip~\cite{Salem:2009}. A time lens based temporal imaging system can be used to temporally magnify the mixed optical signal from a PDV and the velocity range can be extended and/or the bandwidth of readout electronics can be reduced. Notably, temporal magnification with a time lens is comparable in function to the time-stretch technique applied in time-stretch PDV. However, temporal imaging is more efficient in its use of the available optical bandwidth than time stretch techniques and is better suited to directly operate on the optically encoded PDV signals resulting in a simpler hardware configuration. In this paper, first we will explain the concepts of PDV and temporal imaging. Then we will assess the operation of a hardware system through simulations for magnifying PDV signals with a FWM time lens, which will validate the proposed idea of the time-lens PDV (TL-PDV). One of our main findings is that a time-lens with a magnification factor $M$ is feasible for PDV applications. 

\section{Photon Doppler Velocimetry}
\begin{figure}
\vspace{-0.4cm}
  \begin{center}
    \includegraphics[width=0.5\textwidth]{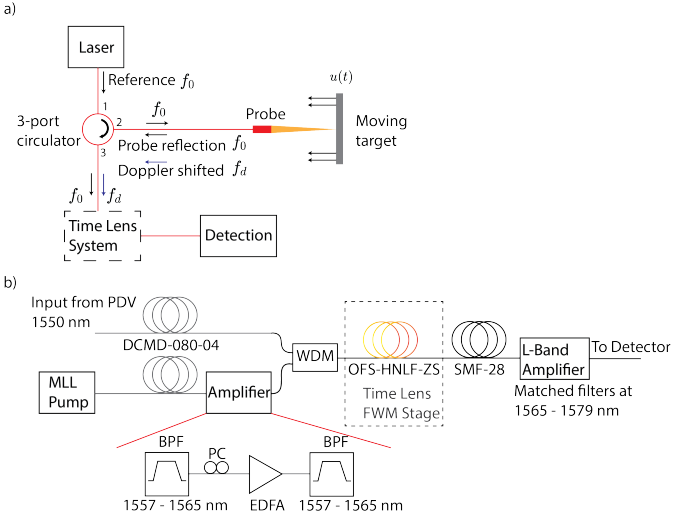}
  \end{center}
  \caption{a) A photon Doppler velocimeter with a time-lens system. Optical fibers are used to transport light from the laser to a probe that launches light to the moving surface. The mixed signals of the reference light and the reflected light from the moving surface result in the signals of the beat frequency and detected~\cite{Strand:2006}. b) A time-lens system can be used to temporally magnify the beat signal with proper fiber lengths for dispersion control. MLL pump: mode-locked laser, BPF: band pass filter, PC: polarization controller, EDFA: Erbium-doped fiber amplifier, WDM: wavelength-division multiplexing, DCMD-080-04 and SMF-28: the model number of Corning optical fiber, and OFS-HNLS-ZS: OFS highly non-linear zero slope optical fiber.}
  \label{fig:pdv}
\end{figure}

We assume that a standard laser source for a PDV system operates in the telecommunications C-band centered at 1550 nm. Notably this is also the wavelength band that is generally leveraged for FWM time-lens systems due to the wide availability of optical fibers, optical amplifiers, and high-performance photodetectors in this range. Figure~\ref{fig:pdv} illustrates a time-lens modified PDV system. The probe focuses the laser light onto the moving surface, and collects the reflected light from it. The frequency ($f_d(t)$) of the reflected light is shifted because of the Doppler effect. The probe is also designed to directly reflect a small amount of non-Doppler-shifted reference light of the frequency ($f_0$) before the moving surface, innately mixing the reference light with the light reflected from the moving surface. This mixed signal returns to the three-port circulator and the mixed signals isolated onto port 3 of the circulator are sent to a photodetector to record the beat signal on a digitizer. The time dependent intensity $I(t)$ that the photodetector measures is given by
\begin{align}
    I(t) = I_0+I_d+\sqrt{I_0 I_d}\sin{(2 \pi f_b(t) t+\phi)}
\end{align}
where $I_0$ is the intensity of the non-Doppler-shifted reference, $I_d$ is the intensity of the Doppler-shifted signal from the moving surface which may vary in time, $f_b$ is the beat frequency, and $\phi$ is the relative phase between the Doppler-shifted and non-Doppler-shifted light. The frequency of the beat signal is defined by the absolute difference between the Doppler-shifted frequency and the non-Doppler-shifted frequency $f_b(t) = |f_d(t)-f_0|$ and can be related to the velocity of the moving surface $v(t)$ by
\begin{align}
    f_b (t) = \pm 2 [\frac{v(t)}{c}]f_0
    \label{eq:fb}
\end{align}
where $c$ is the speed of light. This relation can be derived using the Doppler effect~\cite{Ives:1940} or the geometrical optics with the special relativity for the light reflected from a uniformly moving plane mirror~\cite{Gjurchinovski_2005,Gjurchinovski_2008}. As a point of reference, using Eq.~\ref{eq:fb} with a laser wavelength of 1550 nm ($f_0=$194 THz), a 1 km/s surface velocity corresponds to 1.29 GHz beat frequency. In addition to the single-channel PDV described above, there are PDV configurations using multiple channels, shifted reference frequency, two fibers, etc., which have different stability requirements, advantages, and disadvantages~\cite{Dolan:2020}. For simplicity, in this paper, we only consider the time lens applied to the single-channel PDV system illustrated in Fig.~\ref{fig:pdv}. However, in principle, this time-lens approach can also be applied to other configurations. 

\section{Temporal Imaging}
\begin{figure}[h]
\includegraphics[width=0.45\textwidth]{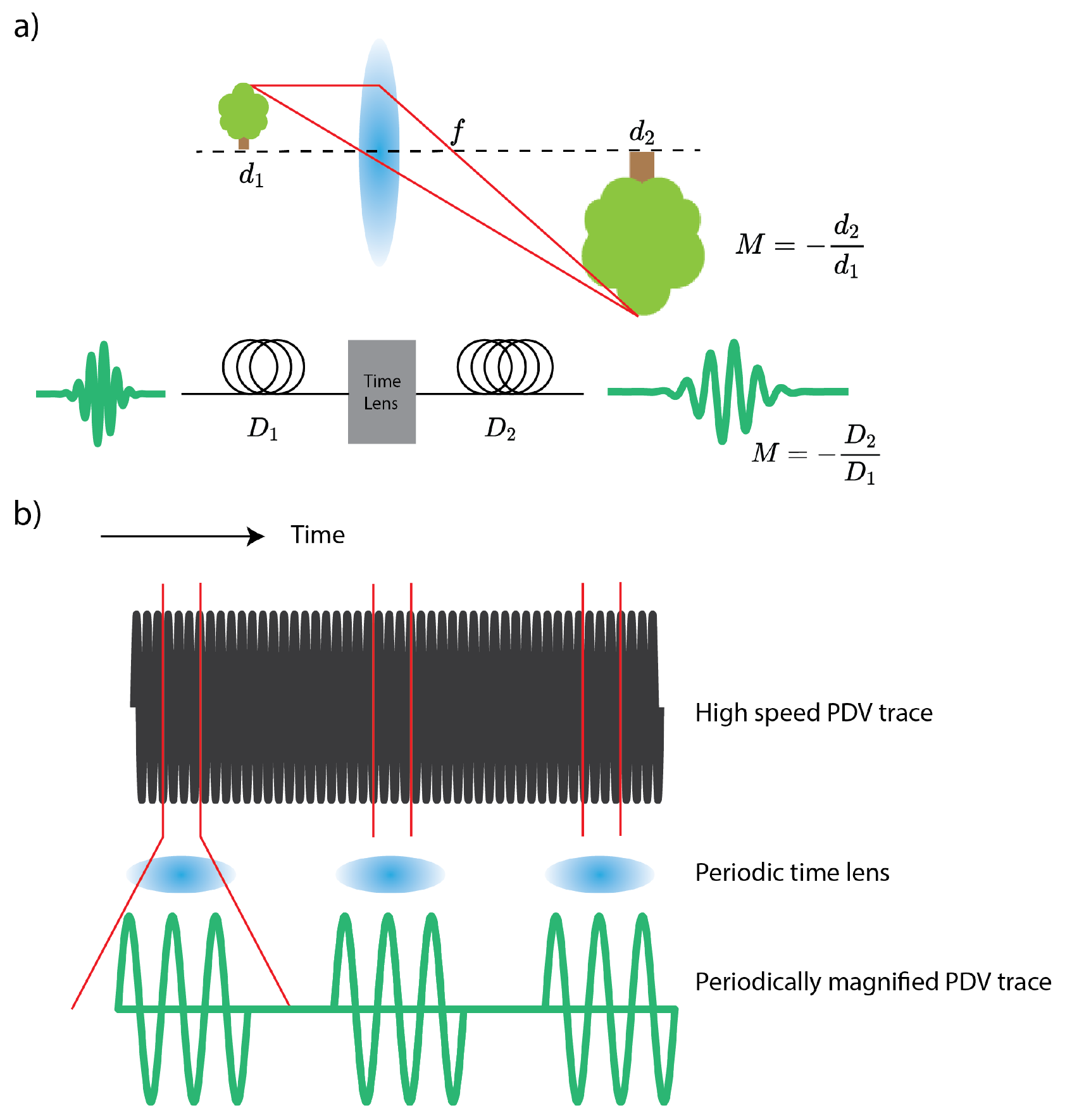}
\caption{a) Space-time imaging analogy. b) Proposed time-lens system can magnify uniformly spaced time windows determined by the pump laser pulse repetition rate.}
\label{fig:timelens_Schematic}
\end{figure}

The concept of a time lens was first demonstrated by Kolner and Nazarathy in 1989~\cite{Kolner:1989} and is rooted in the space-time duality of electromagnetic waves. Figure~\ref{fig:timelens_Schematic} illustrates this duality through the analogous behavior of a time lens and a spatial lens. For example, a spatial lens can generate a magnified spatial image, a time lens can expand a temporal signal in time creating a time magnified temporal signal. The mathematical equivalence between spatial and temporal propagation of light can be proved by using the similarity between the paraxial diffraction in space and dispersion in time~\cite{Kolner:1994}. In the following, we will describe the process of temporal magnification with a time lens, which is naturally divided into three sections: input dispersion, temporal phase modulation, and output dispersion. 

\subsection{Dispersion}
\label{sec:dispersion}
Dispersion is critical to temporal imaging systems as it replicates the effects of spatial diffraction but on temporal signals. In order to mathematically describe the effects of dispersion on a propagating temporal optical waveform, we can consider one-dimensional (1-D) propagation of an optical wave (e.g. plane wave, optical fiber mode, etc.) along the $z$ axis, 
\begin{align}
    A(z,t) = a_0(z,t)e^{i(\omega t -k z)}
\end{align}
where $a_0(z,t)$ is the amplitude and $k = k(\omega)$ is the wavenumber which is a function of oscillation frequency, $\omega$, a characteristic of the material dispersion. Similar to the paraxial wave equation, the propagation of this 1-D waveform through a dispersive medium, such as an optical fiber, can be described by the parabolic equation~\cite{Haus:1984},
\begin{align}
    \frac{\partial^2 a_0}{\partial \tau^2}+\frac{2i}{\frac{\partial^2k}{\partial\omega^2}}\frac{\partial a_0}{\partial \xi}= 0
    \label{eq:parabolic}
\end{align}
where we use the traveling-wave coordinates, i.e., $\tau = (t-t_0)-\frac{(z-z_0)}{v_g}$ and $\xi = z-z_0$ ($t_0$ and $v_0$ are arbitrary constants, and $v_g = \partial\omega/\partial k$ is the group velocity). The general solution of Eq.~\ref{eq:parabolic} is
\begin{align}
    a_0(\xi,\tau) = \frac{1}{2\pi}\int_{-\infty}^{\infty} d\omega ~\hat{a}_0(0,\omega)e^{-\frac{i\xi}{2}(\frac{\partial^2k}{\partial\omega^2})\omega^2}e^{i\omega\tau}
\end{align}
where $\hat{a}_0(0,\omega)$ is the initial envelope spectrum at $\xi=0$.

\subsection{Temporal Phase Modulation and Four-Wave Mixing}
Similar to a spatial lens which can be described by quadratic phase modulation applied in real space~\cite{Goodman:1968}, the effect of the time lens is to impart the waveform $a_0(\xi,\tau)$ with a phase factor that is quadratic in time. For simplicity, it can be realized by an electro-optical phase modulator~\cite{Kolner:1988} driven with a sinusoidal voltage of angular frequency $\omega_m$ when looking locally at the peaks and valleys of the sinusoidal drive signal,
\begin{align}
    e^{i\Gamma_0\cos{(\omega_m \tau+\phi_0)}}\approx  e^{\pm i\Gamma_0(1-\frac{\omega_m^2\tau^2}{2})}\equiv e^{i\varphi_f(\tau)}
\end{align}
where $\Gamma_0$ is the phase modulation amplitude, $\phi_0$ is an adjustable phase difference between the sinusoidal drive signal and the temporal waveform, the $\pm$ accounts for the direction of the local phase shift which can be absorbed into the constant $e^{i\Gamma_0}$. Importantly, since the phase is quadratic in time, the instantaneous frequency has a linear chirp, and the chirp rate is given by $\partial^2\varphi_f/\partial \tau^2=\Gamma_0\omega_m^2$. 

Unfortunately, electro-optical phase modulators are limited in the maximum voltage of the drive signal and therefore cannot generate large phase shifts, limiting the power of the resulting time lens. To address this limitation, parametric nonlinear optical processes have been used to impart a linear frequency chirp and thus quadratic temporal phase shift to an optical signal. FWM time lens can be implemented using materials that lack inversion symmetry such as amorphous silicon waveguides. Wave mixing with nonlinear optical processes can be used to impart extremely large equivalent phase shifts and therefore high-power time lenses can extend the temporal imaging techniques to the sub-picosecond region. Recently, FWM~\cite{Salem:2008} has been used to produce a time lens due to its compatibility with telecommunications wavelengths and fiber optics. As the name implies, FWM is a nonlinear process involving four waves mixing in the medium. Often two of the waves are degenerate and are termed the pump. The pump mixes with a third input wave (the signal) to produce a new wave at a shifted frequency (the idler). Assuming the phase-matching condition has been met, the frequency relation between four waves becomes
\begin{align}
    f_2 = 2f_{p}-f_1,
\end{align}
where $f_{p}$, $f_1$, $f_2$ are the frequencies of the pump, the signal and the idler photons, respectively. 

To mathematically describe the action of the FWM time lens, we can consider a Gaussian pump pulse propagating through the optical fiber and experiencing dispersion as described in Sec.~\ref{sec:dispersion}. Therefore the pulse undergoes temporal dispersive broadening, which results in a linear chirp. This chirped pump pulse undergoes FWM with the input signal (signal, $f_1$) producing the corresponding output signal (idler, $f_2$), which will be modulated with quadratic phase~\cite{Salem:2008}. A time lens can generate a quadratic phase $ \varphi_f(\tau)=\frac{\tau^2}{2\phi_f^{"}}=-\frac{\tau^2}{\phi_p^{"}}$ to the signal where $\phi_{f}^{"} = 1/(\partial^2 \varphi_f/\partial \tau^2)$ is the focal group-delay dispersion (GDD) associated with the lens. $\phi_p^{"} = k_{2}^{(p)}\xi_{p}$ is the GDD experienced by the pump where $k_2^{(p)} = \partial^2 k^{(p)}/\partial \omega^2$ is the group-velocity dispersion (GVD) and $\xi_{p}$ is the length of the dispersive optical fiber. We can describe the dispersion before and after the time lens by their GDD parameters $\phi_1^{"}=k_2^{(1)}\xi_1$ and $\phi_2^{"}=k_2^{(2)}\xi_2$
where $k_2^{(1,2)} = \partial^2 k^{(1,2)}/\partial \omega^2$ are the GVD and $\xi_{1,2}$ are the length of the dispersive optical fibers, respectively. Analogous to the object and image positions in spatial imaging, the GDD relation before and after the time lens can be described by
\begin{align}
    \frac{1}{\phi_1^{"}}+\frac{1}{\phi_2^{"}}=-\frac{1}{\phi_f^{"}}
    \label{eq:focallength}
\end{align}
where the magnification~\cite{Salem:2008} is given by $M=-\phi_2^{"}/\phi_1^{"}$. 

\subsection{Output Waveform from the Temporal Imaging System}
For a single lens imaging process such as the temporal magnification considered here, the temporal imaging system is formed through a combination of the input-dispersion, the quadratic temporal phase modulation (time lens) through FWM, and output-dispersion as shown in Fig.~\ref{fig:timelens_Schematic}. To illustrate this process we consider an arbitrary waveform, $\hat{A}(0,\omega)$, entering the temporal imaging system. The first step is input-dispersion:
\begin{align}
    A(\xi_1,\tau)=&F.T.^{-1}[\hat{A}(0,\omega)\hat{G}_1(\xi_1,\omega)]\notag\\
    =&\frac{1}{2\pi}\int_{-\infty}^{\infty}\hat{A}(0,\omega)e^{-\frac{i\xi_1}{2}(\frac{d^2k_1}{d\omega^2})\omega^2} e^{i\omega\tau}d\omega
\end{align}
where $\hat{G}_1(\xi_1,\omega)=e^{-\frac{i\xi_1}{2}(\frac{d^2k_1}{d\omega^2})\omega^2}$ is the input-dispersion and $F.T.^{-1}$ means the inverse Fourier transformation. Then the second step is the time lens:
\begin{align}
    A(\xi_1+\epsilon,\tau)=&A(\xi_1,\tau)H(\tau)
\end{align}
where $H(\tau)=e^{i\frac{\tau^2}{ 2\phi^{"}_f}}$ is the phase modulation.  Again, we calculate it in the frequency domain:
\begin{align}
    &F.T.^{-1}[\hat{A}(0,\omega)\hat{G}_1(\xi_1,\omega)]\star F.T.^{-1}[\hat{H}(\omega)]\notag\\
    =&F.T.^{-1}[\frac{1}{2\pi}\int_{-\infty}^{\infty}d\omega^{\prime}\hat{A}(0,\omega^\prime)\hat{G}_1(\xi_1,\omega^\prime)\hat{H}(\omega-\omega^\prime)]
\end{align}
where $ \hat{H}(\omega-\omega^\prime)=\sqrt{i2\pi \phi^{"}_f}e^{-i\frac{\phi^{"}_f}{2}(\omega-\omega^\prime)^2}$. The final step is output-dispersion:
\begin{align}
    A(\xi_2,\tau)=&\frac{1}{2\pi}F.T.^{-1}[\hat{A}(0,\omega)\hat{G}_1(\xi_1,\omega)\hat{H}(\omega)\hat{G}_2(\xi_2,\omega)]
\end{align}
which is equal to
\begin{align}
    A(\xi_2,\tau)&=\sqrt{\frac{1}{M}}e^{i\frac{\tau^2}{4(b+c)}}\frac{1}{2\pi}\notag\\
    &\times\int_{-\infty}^{\infty} d\omega \hat{A}(0,\omega)e^{-i(a+\frac{1}{\frac{1}{b}+\frac{1}{c}}){\omega}^2}e^{i\frac{\omega}{M}\tau}
\end{align}
where $a\equiv\frac{\xi_1}{2}(\frac{\partial^2k_1}{\partial{\omega}^2})=\frac{\phi^{"}_1}{2}$, $b\equiv\frac{\xi_2}{2}(\frac{\partial^2k_2}{\partial\omega^2})=\frac{\phi^{"}_2}{2}$, $c\equiv\frac{\phi^{"}_f}{2}$ and $c/(b+c)=1/M$.
\begin{figure}[t]
\includegraphics[width=0.45\textwidth]{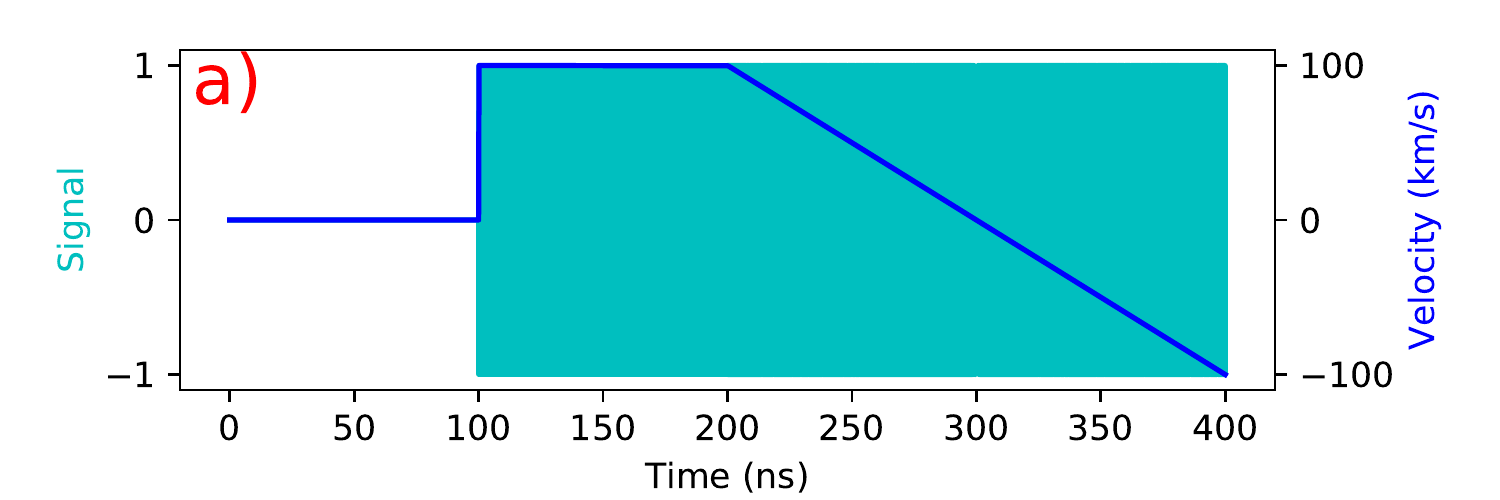}
\includegraphics[width=0.45\textwidth]{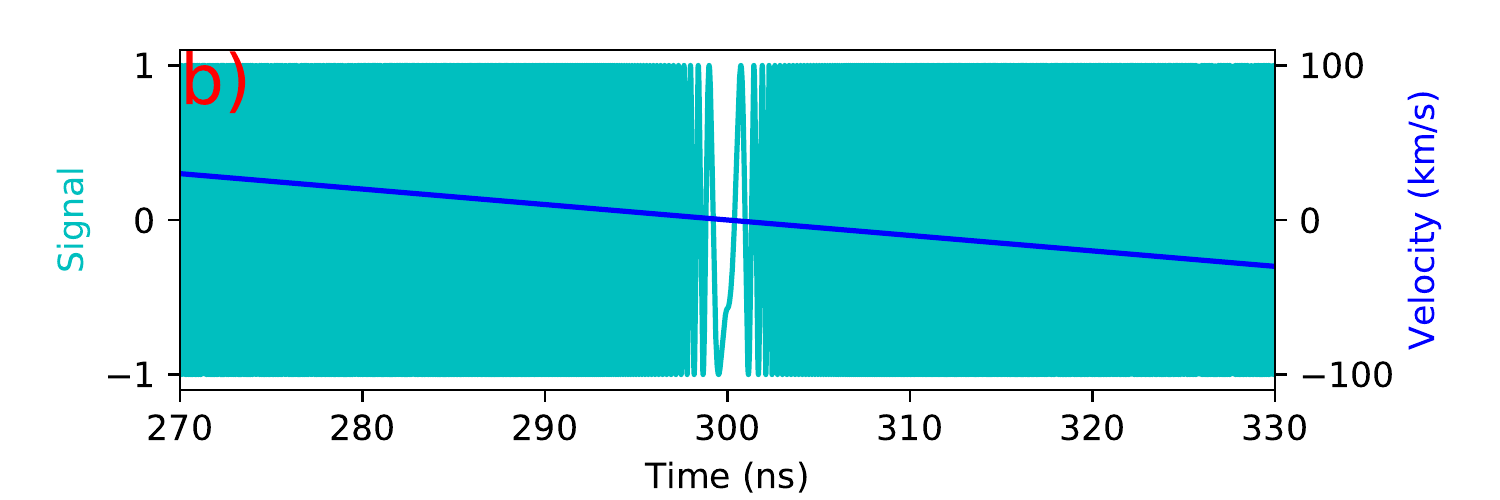}
\includegraphics[width=0.45\textwidth]{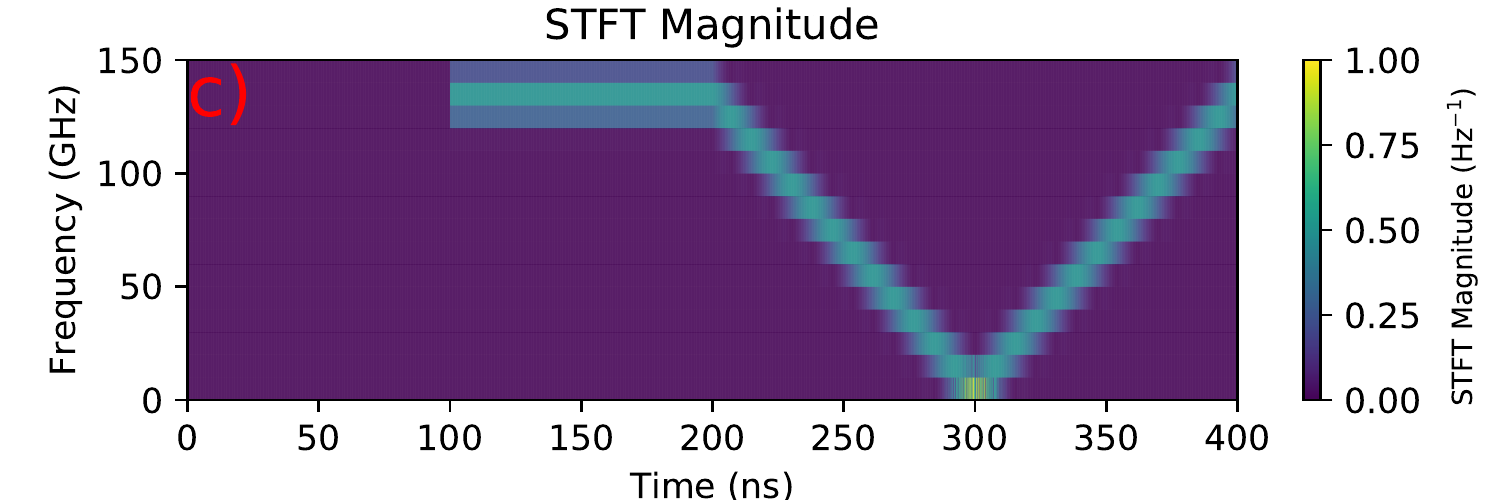}
\caption{a) PDV simulation signals with laser wavelength of 1550 nm. b) Zoomed-in PDV simulation signals when the frequency is around 0. c) Short time Fourier transform of the PDV simulation signal.}
\label{fig:pdv_sim}
\end{figure}
\begin{figure}[t]
\includegraphics[width=0.45\textwidth]{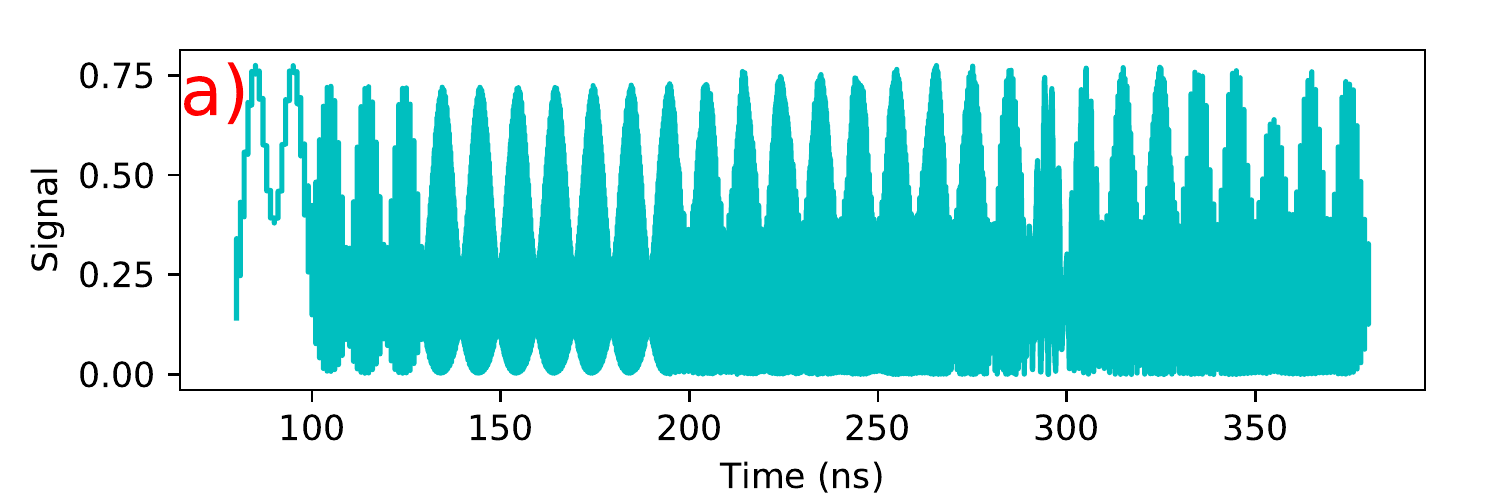}
\includegraphics[width=0.46\textwidth]{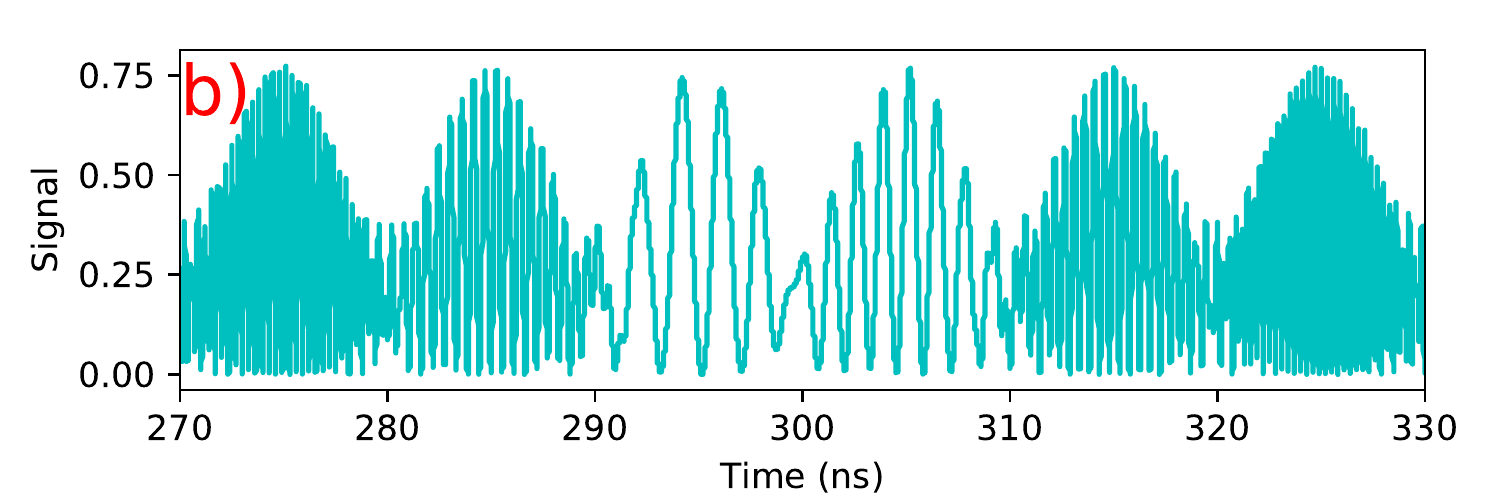}
\includegraphics[width=0.45\textwidth]{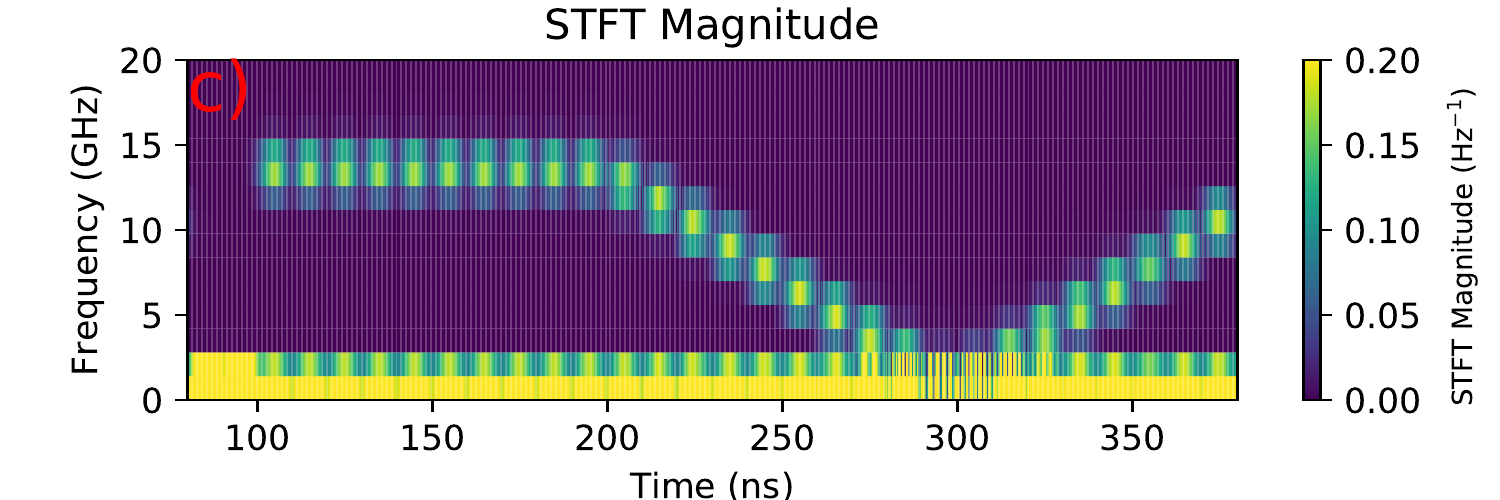}
\caption{a) The PDV simulation signal after the time lens. b) Zoomed-in PDV simulation signal after the time lens. c) Short time Fourier transform of the PDV simulation signal after the time lens. The frequency is significantly reduced (by a factor of 10).}
\label{fig:pdv_sim_timelens}
\end{figure}

\section{Simulation of Time-lens Photon Doppler Velocimetry (TL-PDV)}
The velocity range that can be recorded by a PDV system is ultimately constrained by the electronic bandwidth of the photodetector and the digitizer. A time magnification system, similar to the one in Ref.~\cite{Salem:2009}, can be applied to locally expand the temporal optical waveform, slowing down the beat frequency of PDV and allowing a detection of higher beat frequencies with lower bandwidth electronics. For reference, the current limit of the velocity of PDV is about 50 km/s, which corresponds to a beat frequency of $f_b = 64$ GHz for $f_0 = 194$ THz ($\lambda = 1550$ nm).

In order to demonstrate temporal magnification of PDV signals, we consider an example using a laser of $\lambda = 1550$ nm and the velocity changes from 100 km/s to -100 km/s (the blue curve) similar to the example used in Ref.~\cite{Dolan:2020} and the frequency range is about 120 GHz as shown in Fig.~\ref{fig:pdv_sim}, which would not be measurable by a normal PDV system. In order to simulate the time-lens system, we assume ideal FWM process operating without any bandwidth limitations. Such an assumption is reasonable here since FWM has been demonstrated over bandwidths much larger than those considered here~\cite{Turner-Foster:2010}. For the pump pulse we assume a width of $0.5$ ps and a repetition rate of $100$ MHz. The pump pulse is sent through a third order dispersion (TOD) limited dispersion compensating fiber (i.e. Corning model: DCMD-080-04) of length $\xi_p=\xi/2$ where $\xi=4163$ m. The group velocity dispersion parameter is assumed to be $D=-(2\pi c/\lambda^2)k_2 = 17 \, \text{ps} \, \text{nm}^{-1} \text{km}^{-1}$ for all fibers. Before the time lens, the signal is chirped using the same type of TOD limited fiber with length $\xi_1 = 1.1 \xi$. After the time lens, idler is sent through standard SMF (single mode fiber: e.g. SMF-28) with length $\xi_2 = 11 \xi$ which results in a magnification of $M=10$. The new beat frequency and the velocity equation will be 
\begin{align}
    f_b (t) = \pm\frac{2}{M} [\frac{v(t)}{c}]f_0.
\end{align} As shown in Fig.~\ref{fig:pdv_sim_timelens}, the frequency range will be reduced down to 12 GHz, which is a much smaller frequency for digitization. Notably, the PDV signals are magnified periodically, which corresponds to the pulse repetition rate of the pump laser. For observing dynamics at different time scales, this rate can be engineered to properly sample the velocities by temporally multiplexing or blocking pulses. Also, larger magnification values can be achieved by using larger values of output dispersion. However, care should be taken not to overlap adjacent pulses. Further, for longer propagation lengths, third order dispersion might become an issue and hence would require additional post-processing.
\section{Conclusion}
In the paper, we describe a TL-PDV system using a time lens. As shown in our simulations, a time lens can be used to directly and efficiently extend the velocity range of PDV and allow for the capture of velocity swings in the range $\pm 100$ km/s.
\section{Acknowledgements}
This work is supported in part by DOE/NNSA Office of Experiment Science C3 program. This work is also sponsored in part by the Department of the Defense, Defense Threat Reduction Agency under award HDTRA1-20-2-0001. The content of the information does not necessarily reflect the position or the policy of the federal government, and no official endorsement should be inferred.

The data that support the findings of this study are available from the corresponding author upon reasonable request.

\bibliography{main}

\end{document}